\documentclass[%
 reprint,
 twocolumn,
superscriptaddress,
showpacs,preprintnumbers,
 amsmath,amssymb,
 aps,
prl,
longbibliography,
]{revtex4-1}

\usepackage{tabularx}
\usepackage{multirow}

\usepackage{graphicx}
\usepackage{subfigure}
\usepackage{dcolumn}
\usepackage{bm}
\usepackage{notes2bib} 
\usepackage[colorlinks = true,
            linkcolor = red,
            urlcolor  = blue,
            citecolor = red,
            anchorcolor = blue]{hyperref}

\usepackage[symbol*]{footmisc}
\DefineFNsymbolsTM{myfnsymbols}{%
  \textdagger 
  \textasteriskcentered *
  \dagger
  \textdaggerdbl \ddagger
  \textsection   \mathsection
  \textbardbl    \|%
  \textparagraph \mathparagraph
}%
\setfnsymbol{myfnsymbols}

\begin{document}


\title{First Measurement of Inclusive Muon Neutrino Charged Current \\ Differential Cross Sections on Argon at $E_\nu \sim 0.8$ GeV with the MicroBooNE Detector}


\newcommand{\Bern}{Universit{\"a}t Bern, Bern CH-3012, Switzerland}
\newcommand{\BNL}{Brookhaven National Laboratory (BNL), Upton, NY, 11973, USA}
\newcommand{\Cambridge}{University of Cambridge, Cambridge CB3 0HE, United Kingdom}
\newcommand{\Chicago}{University of Chicago, Chicago, IL, 60637, USA}
\newcommand{\Cincinnati}{University of Cincinnati, Cincinnati, OH, 45221, USA}
\newcommand{\CSU}{Colorado State University, Fort Collins, CO, 80523, USA}
\newcommand{\Columbia}{Columbia University, New York, NY, 10027, USA}
\newcommand{\Davidson}{Davidson College, Davidson, NC, 28035, USA}
\newcommand{\FNAL}{Fermi National Accelerator Laboratory (FNAL), Batavia, IL 60510, USA}
\newcommand{\Harvard}{Harvard University, Cambridge, MA 02138, USA}
\newcommand{\IIT}{Illinois Institute of Technology (IIT), Chicago, IL 60616, USA}
\newcommand{\KSU}{Kansas State University (KSU), Manhattan, KS, 66506, USA}
\newcommand{\Lancaster}{Lancaster University, Lancaster LA1 4YW, United Kingdom}
\newcommand{\LANL}{Los Alamos National Laboratory (LANL), Los Alamos, NM, 87545, USA}
\newcommand{\Manchester}{The University of Manchester, Manchester M13 9PL, United Kingdom}
\newcommand{\MIT}{Massachusetts Institute of Technology (MIT), Cambridge, MA, 02139, USA}
\newcommand{\Michigan}{University of Michigan, Ann Arbor, MI, 48109, USA}
\newcommand{\NMSU}{New Mexico State University (NMSU), Las Cruces, NM, 88003, USA}
\newcommand{\Otterbein}{Otterbein University, Westerville, OH, 43081, USA}
\newcommand{\Oxford}{University of Oxford, Oxford OX1 3RH, United Kingdom}
\newcommand{\PNNL}{Pacific Northwest National Laboratory (PNNL), Richland, WA, 99352, USA}
\newcommand{\Pitt}{University of Pittsburgh, Pittsburgh, PA, 15260, USA}
\newcommand{\StMarys}{Saint Mary's University of Minnesota, Winona, MN, 55987, USA}
\newcommand{\SLAC}{SLAC National Accelerator Laboratory, Menlo Park, CA, 94025, USA}
\newcommand{\SDSMT}{South Dakota School of Mines and Technology (SDSMT), Rapid City, SD, 57701, USA}
\newcommand{\Syracuse}{Syracuse University, Syracuse, NY, 13244, USA}
\newcommand{\TelAviv}{Tel Aviv University, Tel Aviv, Israel, 69978}
\newcommand{\Tennessee}{University of Tennessee, Knoxville, TN, 37996, USA}
\newcommand{\UTA}{University of Texas, Arlington, TX, 76019, USA}
\newcommand{\Tubitak}{TUBITAK Space Technologies Research Institute, METU Campus, TR-06800, Ankara, Turkey}
\newcommand{\Tufts}{Tufts University, Medford, MA, 02155, USA}
\newcommand{\VTech}{Center for Neutrino Physics, Virginia Tech, Blacksburg, VA, 24061, USA}
\newcommand{\Warwick}{University of Warwick, Coventry CV4 7AL, United Kingdom}
\newcommand{\Yale}{Wright Laboratory, Department of Physics, Yale University, New Haven, CT, 06520, USA}

\affiliation{\Bern}
\affiliation{\BNL}
\affiliation{\Cambridge}
\affiliation{\Chicago}
\affiliation{\Cincinnati}
\affiliation{\CSU}
\affiliation{\Columbia}
\affiliation{\Davidson}
\affiliation{\FNAL}
\affiliation{\Harvard}
\affiliation{\IIT}
\affiliation{\KSU}
\affiliation{\Lancaster}
\affiliation{\LANL}
\affiliation{\Manchester}
\affiliation{\MIT}
\affiliation{\Michigan}
\affiliation{\NMSU}
\affiliation{\Otterbein}
\affiliation{\Oxford}
\affiliation{\PNNL}
\affiliation{\Pitt}
\affiliation{\StMarys}
\affiliation{\SLAC}
\affiliation{\SDSMT}
\affiliation{\Syracuse}
\affiliation{\TelAviv}
\affiliation{\Tennessee}
\affiliation{\UTA}
\affiliation{\Tubitak}
\affiliation{\Tufts}
\affiliation{\VTech}
\affiliation{\Warwick}
\affiliation{\Yale}

\author{P.~Abratenko} \affiliation{\Michigan} 
\author{C.~Adams} \affiliation{\Harvard}
\author{M.~Alrashed} \affiliation{\KSU}
\author{R.~An} \affiliation{\IIT}
\author{J.~Anthony} \affiliation{\Cambridge}
\author{J.~Asaadi} \affiliation{\UTA}
\author{A.~Ashkenazi} \affiliation{\MIT}
\author{M.~Auger} \affiliation{\Bern}
\author{S.~Balasubramanian} \affiliation{\Yale}
\author{B.~Baller} \affiliation{\FNAL}
% ...

\collaboration{The MicroBooNE Collaboration} 
\email{microboone\_info@fnal.gov}
\noaffiliation

\date{\today}

\begin{abstract}

\begin{center}
Dedicated to the memory of Simone Marcocci.
\end{center}

We report the first measurement of the double-differential and total muon neutrino charged current inclusive cross sections on argon at a mean neutrino energy of 0.8 GeV.
Data were collected using the MicroBooNE liquid argon time projection chamber located in the Fermilab Booster neutrino beam and correspond to $1.6 \times 10^{20}$ protons on target of exposure. The measured differential cross sections are presented as a function of muon momentum, using multiple Coulomb scattering as a momentum measurement technique, and the muon angle with respect to the beam direction. We compare the measured cross sections to multiple neutrino event generators and find better agreement with those containing more complete treatment of quasielastic scattering processes at low  $Q^2$. The total flux integrated cross section is measured to be $0.693 \pm 0.010 \, (\text{stat}) \pm 0.165 \, (\text{syst}) \times 10^{-38} \, \text{cm}^{2}$.
\end{abstract}

\maketitle


Current and next generation precision neutrino oscillation experiments aim to probe beyond standard model physics, such as $CP$ violation in the lepton sector and sterile neutrinos. These experiments measure the oscillation probability which, in the experimental setup, is convolved with models of the neutrino interaction cross section and kinematics of secondary leptons and hadrons emerging from the neutrino's interaction. This link is complicated by the existence of nuclear effects and final-state interactions, which to date cannot be modeled precisely, in particular for heavy target nuclei typically used in modern neutrino experiments. Many future experiments, including DUNE \cite{dune1, dune2, dune3} and the SBN \cite{sbn} program, employ liquid argon time projection chambers (LArTPCs) as detectors. As a consequence, neutrino-argon cross section measurements have paramount importance, especially given the relative scarcity of neutrino-argon data \cite{ArgoNeuTCCincl, ArgoNeuTCCincl2}.

We present the first $\nu_{\mu}$ charged current (CC) inclusive double-differential (in muon momentum and scattering angle) cross section measurement on argon. Neutrinos in the same $\sim$1 GeV energy range will be studied by the SBND and ICARUS experiments, and this is the energy where DUNE is very sensitive to precise measurements of oscillation parameters as $CP$ violating effects are maximized at this energy and DUNE's baseline. 
The inclusive CC process, in which only the outgoing muon is required to be reconstructed, comprises multiple interaction processes and is dominantly quasielastic scattering in the case of MicroBooNE \cite{zeller}. 
Inclusive measurements are particularly important as the clear signal definition allows a straightforward comparison to theory models and other experiments. They are also the foundation for studies of more complex event topologies involving detection of hadrons in the final state.
With the fully active and high-resolution MicroBooNE LArTPC detector, the outgoing muon phase space can be probed with full acceptance in both angle and momentum for the first time. 
The momentum of the outgoing muon is measured by using multiple Coulomb scattering (MCS) \cite{mcs} thus allowing the analysis sample to include both exiting and contained muons.

%
%
The MicroBooNE detector has 85 tons of liquid argon active mass and is located along the Booster Neutrino Beam (BNB) at Fermilab, 463 m from the target. The BNB consists primarily of $\nu_\mu$ (93.6\%) with energy from a few tens of MeV to $\sim$2~GeV.
If neutrinos interact in the MicroBooNE detector \cite{det}, charged particles traverse a volume of highly pure liquid argon leaving trails of ionization electrons along their paths, and also create prompt ultraviolet scintillation photons.
Ionization electrons drift horizontally and transverse to the neutrino beam direction in an electric field of 273 V/cm to a system of three anode wire planes located 2.5~m from the cathode plane and are detected by electronics immersed in the liquid argon \cite{noise_paper}.
Scintillation photons are observed by 32 photomultipliers (PMTs) \cite{pmt_paper}. 

%
%
The data used in this analysis are taken from an exposure of $1.6 \times 10^{20}$ protons on target (POT), after applying data quality criteria for the beam and detector operating conditions. This corresponds to a four-month exposure, from February to July 2016.
Two different data streams are used in this analysis: an on-beam data sample, triggered by BNB neutrino spills, and an off-beam data sample, taken during periods when no beam was received. The off-beam data sample is used for a measurement of cosmic ray (CR) backgrounds, which is important because the MicroBooNE detector operates on Earth's surface.

%
%
The flux of neutrinos at the MicroBooNE detector is simulated using the framework built by the MiniBooNE Collaboration \cite{miniboone_flux}.
Neutrino interactions in the MicroBooNE detector are simulated using the \textsc{genie} event generator \cite{[{}][{, version 2\_12\_2.}]GENIE}, which generates the primary neutrino interaction inside the nucleus, the production of all final-state particles in the nucleus (hadronization), and the transport and rescattering of the final-state particles through the nucleus (final-state interactions).
CRs crossing the detector volume within the readout window of neutrino events are simulated with \textsc{corsika} \cite{[{}][{, version 7.4003 with constant mass composition model.}]corsika}.
Particle propagation is based on \textsc{geant4}~\cite{geant}, while the simulation of the MicroBooNE detector is performed in the \textsc{larsoft} framework~\cite{larsoft} and includes the generation of wire signals and the modeling of scintillation light in the PMTs.

%
%

Data processing begins with a requirement that PMT activity occurs in coincidence with the arrival of neutrinos. This PMT trigger results in a negligible loss of signal. TPC waveforms originate from drift electrons inducing bipolar signals on the first two wire planes and a unipolar signal on the last plane, which collects the electrons. A noise filtering algorithm removes inherent and electronic noise \cite{noise_paper}, and the signals are deconvolved to a Gaussian to further eliminate detector artifacts \cite{baller}. Individual signal waveforms are identified as hits and are sorted spatially to form clusters. Clusters are matched across planes and identified as tracklike or showerlike by the Pandora multialgorithm pattern recognition framework \cite{pandora}. Optical reconstruction combines correlated PMT waveforms across the detector into 
flashes.

A series of algorithms is used to identify and remove CRs.
These algorithms identify tracks that traverse the detector from top to bottom, adding optical information to identify CRs that enter from the anode or cathode planes. Stopping muon tracks originating outside the detector are identified as CRs either by their Bragg peak or by their Michel decay.

%
%
This analysis makes use of the optical system to reduce the high CR rate by more than 3 orders of magnitude. The observed reconstructed optical neutrino candidate flash, which is the spatial intensity distribution of scintillation light arriving at the PMTs behind the anode plane during the 1.6~$\mu$s beam window, is compared to a light prediction made from TPC tracks and showers originating from a common vertex found in the same $\sim$1~ms duration event. The best matched pair (if any) identifies the neutrino interaction and its secondary tracks and showers within the event. 
Calorimetric information~\cite{calo_paper} in the form of a truncated mean value of the deposited charge per unit length, $dQ/dx$, and track length are used to discriminate muons from protons.
The candidate interaction must contain a track that has a measured $dQ/dx$ compatible with a muon. In cases where there are multiple particles originating from the same vertex (predicted to occur in 70\% of the neutrino events), only one such muonlike track is identified by selecting the longest track. The purity of selecting a true muon is 95\%. Several algorithms ensure the quality of the fitted track by limiting the allowed spatial dispersion of the reconstructed hits with respect to the track hypothesis.

The momentum of the muon is measured using multiple Coulomb scattering. Here, the magnitude of the momentum is a fit parameter that describes the scattering pattern of the track \cite{mcs}. 
The strength of this algorithm is that it can estimate the muon momentum for muon tracks spatially contained in the detector as well as exiting tracks, which is important given that only 30\% of neutrino-induced muons are spatially contained. In addition, the measured momentum using range is used to identify and exclude misreconstructed tracks by comparing it to the MCS momentum. The two momentum estimates would disagree if the reconstructed track is incomplete or inaccurate.

\begin{figure*}[]
  \centering
  \subfigure[]{\includegraphics[width=.45\textwidth]{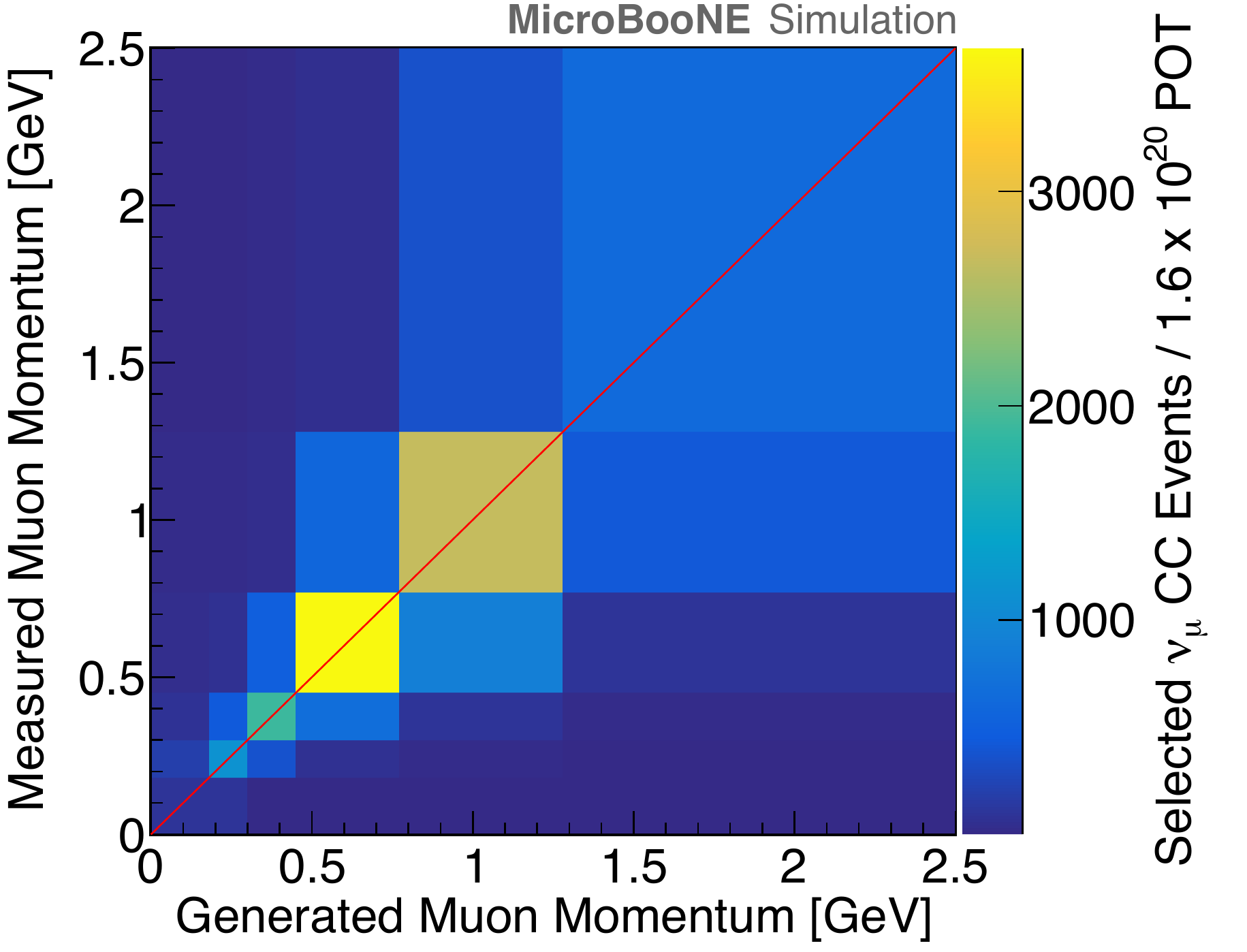}}\quad\quad\quad
  \subfigure[]{\includegraphics[width=.45\textwidth]{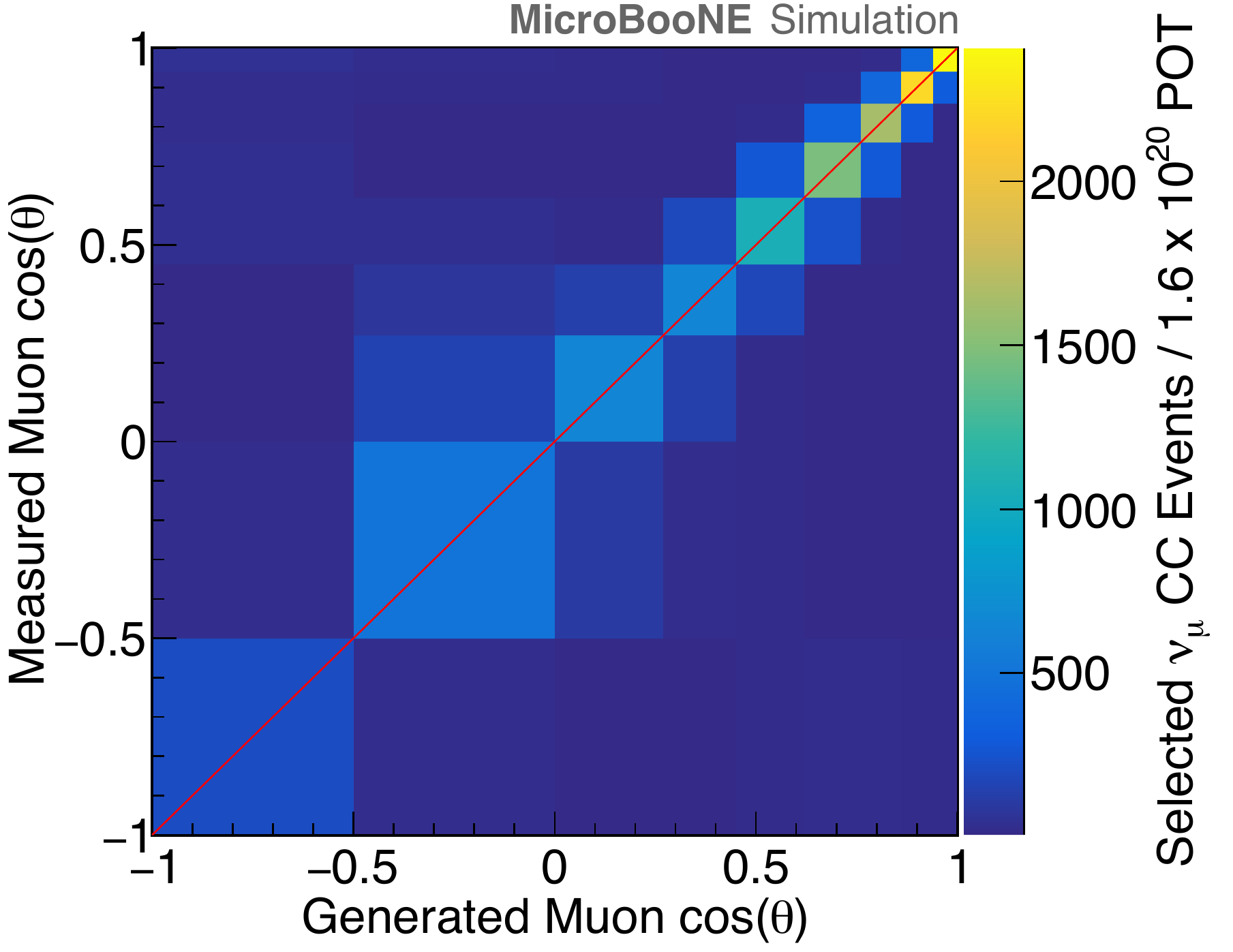}}
  \caption{Comparison of the generated versus measured muon momentum (a) and cosine of the muon angle (b) for the simulated and selected $\nu_\mu$ CC events, showing a 10\%--15\% momentum resolution and few-degree angular resolution. The binning is the same as used in the cross section extraction.}
  \label{fig:binning}
\end{figure*}

Figure~\ref{fig:binning} shows the measured versus generated muon momentum $p_\mu$ and the measured versus generated $\cos\theta_\mu$ for simulated events, where $\theta_\mu$ is the muon angle with respect to the incident neutrino beam direction.
There is a possibility for tracks to be misreconstructed with the opposite direction~\cite{pandora}. The impact is strongest in the two backwards bins, $\cos\theta_\mu \in [-1, -0.5)$ and $\cos\theta_\mu \in [-0.5, 0)$, where only 46\% and 56\% of events come from the same bin they were generated in, respectively. The other events are actually forward going but get reconstructed with the opposite direction~\cite{pandora}. This effect is included in the smearing matrix; therefore, the muon kinematic distributions of data and simulation remain comparable with each other.

The final selected sample contains 27 200 events. The signal selection efficiency, measured in simulation, is 57.2\%. The selection accepts events across the entire angular phase space. The purity of the final selection is 50.4\%. The efficiency and purity are relatively flat as a function of muon kinematics and the number of final-state particles. Different interaction processes have approximately the same efficiency. 
The main backgrounds in this analysis are (i) CRs that overlap in time with the beam spill and trigger the readout (estimated to be 29.1\% of all selected events), (ii) CRs overlaid with neutrino interactions in which the cosmic muon was misidentified as coming from a neutrino interaction (6.4\%), (iii) neutrinos that interact outside the fiducial volume with an entering track selected as the muon candidate (7.6\%), (iv) events in which a neutrino interacts outside the cryostat but the muon enters the TPC and is selected (here called ``dirt'' interactions) (4.4\%), (v) neutral current interactions where a final state particle is misidentified as a muon (1.6\%), (vi) beam intrinsic muon antineutrino interactions (0.4\%), and (vii) beam intrinsic electron (anti)neutrino interactions (0.1\%).
The largest background (i) is measured with off-beam data and subtracted from the on-beam data. The off-beam data sample has twice the statistics of the on-beam data. Other backgrounds are estimated from simulation.
The accuracy of the detector modeling has been verified by looking at selected event distributions of variables not affected by the neutrino interaction physics,  e.g., the interaction points in the detector, where we have good data to simulation agreement.

%
%

This analysis measures the double-differential $\nu_{\mu}$ CC cross section on argon as a function of the muon momentum $p_\mu$ (measured using MCS) and the cosine of the muon angle $\theta_\mu$ with respect to~the beam direction. The flux-integrated, double-differential cross section measured in bin $i$ is defined as
\begin{equation}
\label{eq:xsec_differential}
\left ( \frac{d^2\sigma}{dp_\mu d\cos\theta_\mu} \right )_{i} = \frac{N_{i} - B_{i}}{\tilde{\epsilon}_{i} T \Phi_{\nu_\mu} (\Delta p_\mu \Delta \cos\theta_\mu)_i},
\end{equation}
where $N_i$, $B_i$ and $\tilde{\epsilon_i}$ are the number of selected data events, the expected background events, and the detection efficiency in bin $i$. $(\Delta p_\mu \Delta \cos\theta_\mu)_i$ is the $i\text{th}$ bin area. $T$ and $\Phi_{\nu_\mu}$ are the number of target nucleons, and the integrated BNB muon neutrino flux from 0 to 10 GeV. The total integrated BNB $\nu_\mu$ flux in neutrino mode running, corresponding to $1.6 \times 10^{20}$ POT, is $\Phi_{\nu_\mu} = 1.16 \times 10^{11} \, \nu_\mu / \text{cm}^{2}$, and its mean neutrino energy is 823 MeV. The relevant energy range for this measurement is from 325 to 1325 MeV, which includes 68\% of neutrinos from the BNB.

We report the final cross section result as a function of measured kinematic variables following a ``forward-folding'' approach.
This is done using a migration matrix $S$, which transforms the number of generated events $N^\text{gen}_j$ in a bin $j$ of generated momentum and angle from any inclusive model to the number of events $N_i$ in a bin $i$ of measured momentum and angle. $N_i = \sum_{j=1}^{M} S_{ij} N^\text{gen}_j$, where $S$ is given by $
S_{ij} = P(\text{measured in bin}\, i \,|\, \text{generated in bin}\, j)$
and $M$ is the total number of bins. 

The efficiency correction as a function of the measured quantities $\tilde{\epsilon}_i$ that we applied to our data as described in Eq.~\eqref{eq:xsec_differential} is given by
\begin{equation}
\label{eq:eff_smear}
\tilde{\epsilon}_i = \frac{ \sum_{j=1}^{M} S_{ij}N^\text{sel}_j}{ \sum_{j=1}^{M} S_{ij}N^\text{gen}_j},
\end{equation}
where $N^\text{sel}_j$ is the number of selected signal events in bin $j$.

%
%
The uncertainty on the measurement is dominated by systematic uncertainties, which come from the neutrino flux, neutrino interaction model, and detector response.
Uncertainties, both statistical and systematic, are encoded in a covariance matrix $E$.
The total uncertainty matrix is a combination of the statistical and systematic errors, $E = E^\text{stat} + E^\text{syst}$, where $E^\text{stat}$ is the statistical uncertainty matrix and $E^\text{syst}$ is the systematic covariance matrix. 

\begin{table}
\caption{Contributions to the total cross section systematic uncertainty.}
\label{tab:all_syst}
\begin{ruledtabular}
\begin{tabular}{cc}
Source of uncertainty      & Relative uncertainty  [\%]\\
\hline
Beam flux                  &  12.4                     \\
Cross section modeling     &  3.9                      \\
Detector response          &  16.2                     \\
Dirt background            &  10.9                     \\
Cosmic ray background      &  4.2                      \\
MC statistics              &  0.2                      \\
\hline
Statistics                 &  1.4                      \\
\hline
Total                      &  23.8                     \\
\end{tabular}
\end{ruledtabular}
\end{table}

\begin{figure*}
\includegraphics[width=1.0\textwidth]{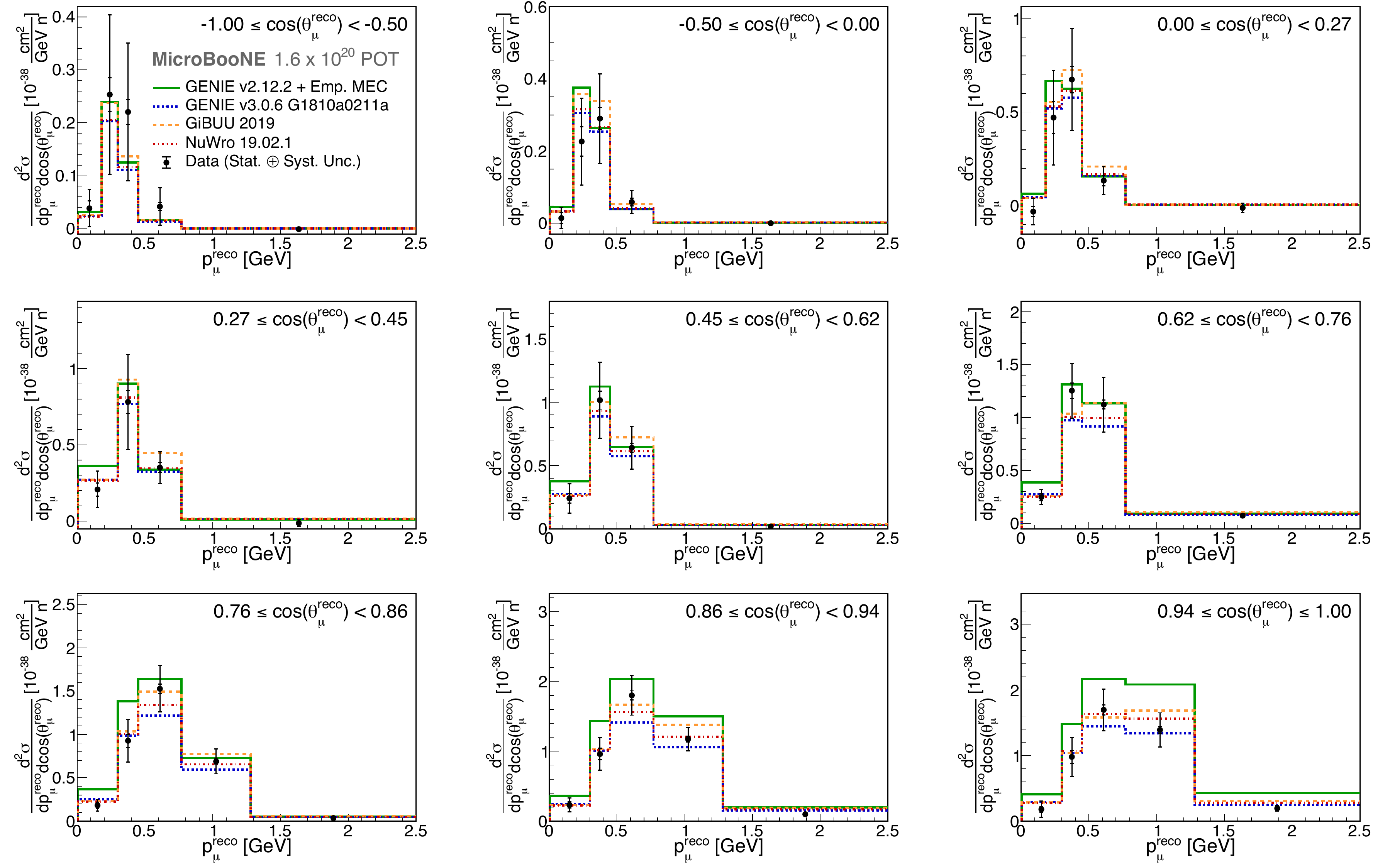}
\caption{$\nu_\mu$ CC inclusive double-differential cross section on argon per nucleon $n$ as a function of the measured muon momentum and cosine of the measured muon polar angle (angle with respect to the incoming neutrino direction), $d^2\sigma/(dp_\mu^\text{reco}d\cos\theta_\mu^\text{reco}) [10^{-38} \text{cm}^2/(\text{GeV} n)]$. 
The data (black) are compared to a \textsc{genie} v2 with empirical MEC prediction (green), a \textsc{genie} v3 prediction (blue), a \textsc{gibuu} prediction (orange), and a \textsc{nuwro} prediction (red), as described in the text. The vertical bars show statistical and systematic uncertainties.}
\label{fig:xsec}
\end{figure*}

To assess the uncertainties on the neutrino flux prediction, the final flux simulation from the MiniBooNE Collaboration is utilized~\cite{miniboone_flux}, updated to the MicroBooNE detector location.
For neutrino cross section modeling uncertainties, we use the \textsc{genie} framework of event reweighting \cite{GENIE, GENIE_reweighting} with its standard reweighting parameters. 
For both cross section and flux systematics we use a \emph{multisim} technique~\cite{roe}, which consists of generating many MC replicas, each one called a ``universe,'' where parameters in the models are varied within their uncertainties. Each universe represents a different reweighting. The simultaneous reweighting of all model parameters allows the correct treatment of correlations among them.
$N$ such universes are then created that can be combined to construct the covariance matrix:
\begin{equation}
\label{eq:multisim}
E_{ij} = \frac{1}{N} \sum_{n = 1}^{N} (\sigma_i^n - \sigma_i^\text{cv})(\sigma_j^n - \sigma_j^\text{cv}),
\end{equation}
where $\sigma$ is a shorthand notation for the double-differential cross section in Eq.~\eqref{eq:xsec_differential}, $i$ and $j$ correspond to bins in measured quantities, $\sigma^\text{cv}_i$ is the central value cross section in bin $i$, and $\sigma_i^n$ is the cross section evaluated in the systematic universe $n$. 

A different model is followed for systematics associated with the detector model, as these systematics are dominated by single detector parameters and are not possible to estimate through reweighting. In this case \emph{unisim} samples~\cite{roe} are generated, where only one detector parameter at a time is changed by its uncertainty. For $M$ detector parameters, the covariance matrix is
\begin{equation}
\label{eq:unisim}
E_{ij} = \sum_{m = 1}^{M} (\sigma_i^m - \sigma_i^\text{cv})(\sigma_j^m - \sigma_j^\text{cv}).
\end{equation}

The total flux, cross section, and detector uncertainties amount to 12\%, 4\%, and 16\% of the total cross section, respectively.
The largest individual contribution to the detector uncertainty comes from using a simple model to simulate the induced charge on neighboring wires of the TPC, leading to a 13\% uncertainty on the total cross section. 
Additional uncertainties are assessed on the dirt and simulated CR background interactions overlaying neutrino interactions, which yield 11\% and 4\% uncertainties on the final cross section measurement, respectively. 
A summary of systematic uncertainty is shown in Table~\ref{tab:all_syst}.

%
%

The double-differential cross section is presented in Fig.~\ref{fig:xsec} and compared with several predictions from different generators. The first uses the default \textsc{genie} configuration in \textsc{genie} v2.12.10, with the addition of a meson exchange current (MEC) interaction channel modeled with an empirical approach~\cite{mec_dytman}. We also compare to the most recent version of \textsc{genie} --- v3.0.6 --- in which we use the {G18\_10a\_02\_11a} comprehensive model configuration.  This includes a number of theoretically motivated improvements. It replaces the Bodek-Ritchie Fermi gas nuclear model with a local Fermi gas (LFG) for the nuclear initial state. The Valencia model is used for quasielastic and MEC interactions~\cite{nieves, nieves2}, and the Kuzmin-Lyubushkin-Naumov~\cite{kuzmin} and Berger-Seghal~\cite{berger_sehgal} model with form factors from MiniBooNE data~\cite{mb_res} for resonant pion production. We also compare to predictions from \textsc{nuwro} and \textsc{gibuu}. \textsc{nuwro}~19.02.1~\cite{nuwro} uses a similar set of models to the \textsc{genie}~v3.0.6 configuration, though the resonant pion production form factors are modified~\cite{graczyk} and the final-state interaction model is the Oset intranuclear cascade model~\cite{salcedo}. \textsc{gibuu}~{2019}~\cite{gibuu} has consistent nuclear medium corrections throughout. It also uses a local Fermi gas model to describe the nucleon momenta, a separate MEC model~\cite{gibuu_mec}, and propagates final state particles according to the Boltzmann-Uehling-Uhlenbeck transport equations.

This is the first test of neutrino event generators against double-differential neutrino scattering data on argon. As is also seen in comparisons to neutrino data on carbon \cite{Ruterbories:2018gub, Abe:2018uhf}, high $\chi^2$ values between data and predictions are observed taking into account the full covariance matrix with off-diagonal elements (not displayed in Fig.~\ref{fig:xsec}).
The largest disagreements between the data and predictions are observed in the high-momentum bins in the most forward-going muon angular bins of $0.94 \leq \cos\theta_\mu \leq 1$ and $0.86 \leq \cos\theta_\mu < 0.94$. 
This region strongly disfavors the \textsc{genie} v2 with empirical MEC.  Other predictions show less tension with the data in this phase space with the exception of the highest momentum bin with the angular range of $0.86 \leq \cos\theta_\mu < 0.94$.
The lowest $\chi^2$ value is obtained for the \textsc{genie} v3 model with a $\chi^2$ of 103.9 for 42 bins. The reduced tension originates from the overall reduced cross section in the forward direction when adopting the local Fermi gas nuclear initial state, which is expected  to be a more realistic momentum distribution of the initial state nucleons, and to a lesser extent the random phase approximation (RPA) correction as included in the \textsc{genie} v3 and \textsc{nuwro} predictions. These effects have the largest impact at low neutrino energies and for heavy nuclear targets, which explains why MicroBooNE is more sensitive to these effects than previous experiments. A perfect backscattered CCQE muon at 800 (1325) MeV neutrino energy corresponds to the square of the four-momentum transfer $Q^2$ of 2.5 (7.0) GeV$^2$. For this reason, these new MicroBooNE cross section results are very valuable for addressing and testing the details of the current neutrino cross section models and for making progress in understanding the physics associated with neutrino interactions. Smearing and covariance matrices, as well as cross section and flux tabulated values, are available in the Supplemental Materials~\cite{suppl}.

Additionally, we compute a flux-integrated cross section $\sigma(\nu_\mu + \text{Ar} \rightarrow \mu^- + X)$ per nucleon of
\begin{equation}
\sigma = 0.693 
         \pm 0.010 \, \text{(stat)} 
         \pm 0.165 \, \text{(syst)} 
         \times 10^{-38} \,\text{cm}^2,
\end{equation}
which is obtained by integrating the number of signal and background events, as well as the efficiency over all bins. 
The measured flux-integrated cross section agrees with the predictions from the models described above within  uncertainty, with \textsc{genie} v2 giving the largest discrepancy.

%
%

In summary, we have reported the first double-differential $\nu_\mu$ charged current inclusive cross section on argon. The presented analysis has full angular coverage and uses multiple Coulomb scattering to estimate the muon momentum, a significant step forward for the LArTPC technology.

As shown in the comparison with various predictions, these data provide a way to differentiate models in neutrino event generators. These measurements not only inform the theory of neutrino nucleus scattering, but also reduce the systematic uncertainties associated with cross section measurements in neutrino oscillation experiments.

This document was prepared by MicroBooNE using the resources of the Fermi National Accelerator Laboratory (Fermilab), a U.S. Department of Energy, Office of Science, HEP User Facility. Fermilab is managed by Fermi Research Alliance, LLC (FRA), acting under Contract No. DE-AC02-07CH11359.
MicroBooNE is supported by the U.S. Department of Energy, Office of Science, Offices of High Energy Physics and Nuclear Physics; the U.S. National Science Foundation; the Swiss National Science Foundation; the Science and Technology Facilities Council (STFC), part of UK Research and Innovation; and The Royal Society (United Kingdom). Additional support for the laser calibration system and CR tagger was provided by the Albert Einstein Center for Fundamental Physics, Bern, Switzerland.


\bibliography{ccinc_ub_prl}

\end{document}